\documentclass{epl}

\usepackage{graphics}
\usepackage{psfig}

\newcommand{\lav}{\langle\,}
\newcommand{\rav}{\,\rangle}
\newcommand{\bem}{\begin{em}}
\newcommand{\eem}{\end{em}}
\newcommand{\beq}{\begin{equation}}
\newcommand{\eeq}{\end{equation}}
\newcommand{\eve}{\vect{e}}
\newcommand{\rve}{\vect{r}}

\title{Are there localized saddles behind the heterogeneous dynamics
  of supercooled liquids?}
\shorttitle{Are there localized saddles behind...}
\author{D. Coslovich\inst{1}\thanks{Email: \email{coslo@ts.infn.it}} 
  \and G. Pastore\inst{1,2}\thanks{Email: \email{pastore@ts.infn.it}}}

\institute{                    
  \inst{1} Dipartimento di Fisica Teorica, Universit{\`a} di Trieste - 
  Strada Costiera 11, 34100 Trieste, Italy\\
  \inst{2} CNR-INFM Democritos National Simulation Center -
  Via Beirut 2-4, 34014 Trieste, Italy
}


\begin{document}

\maketitle

\begin{abstract}
We numerically study the interplay between heterogeneous dynamics and 
properties of negatively curved regions
of the potential energy surface in a model glassy system.
We find that the unstable modes of saddles and quasi-saddles undergo a localization 
transition 
close to the Mode-Coupling critical temperature. 
We also find evidence of a positive 
spatial correlation between clusters of particles having large displacements 
in the unstable modes and dynamical heterogeneities.
\end{abstract}

The dynamics of supercooled liquids is often described as a
complex trajectory across their Potential Energy Surface (PES)\cite{art:sciortino05}. This
approach traces its origin back to the pioneering work of Goldstein
\cite{art:goldstein}, who argued that at low temperature these systems get
trapped in the basins of attraction of local minima of the PES and
hence the atomic dynamics is slowed down.
In the last decade, several authors have addressed a quantitative
study of the PES through numerical simulations of model systems.
The original picture has evolved into a refined description of the dynamics
in terms of collections of local
minima (\bem metabasins \eem~\cite{art:stillinger})
and transitions between them
\cite{art:doliwa03a,art:doliwa03b,art:doliwa03}.
 
Other works have focused
on the properties of the Hessian matrix of the potential
energy, revealing even a more complex scenario, where not only local minima but also 
higher order stationary points (saddles) and more general points characterizing 
regions of negative curvature of PES (quasi-saddles) play an important role in the 
structural slowing down of the liquid
\cite{art:sw82,art:sastry98,art:schroeder,art:broderix,art:grigera02,
art:angelani02,art:sciortino03,art:doliwa03a,art:doliwa03b,art:doliwa03,art:doye02,art:wales03}.
Indeed, it has been shown~\cite{art:angelani02,art:sciortino03} that
some information about the liquid-like diffusive dynamics is  encoded in the imaginary
spectrum of the Hessian matrix: the number of unstable modes $n_{im}$ of saddles is 
correlated with the diffusivity of supercooled model
systems  and decreases as the liquid is cooled. Attempts have been made
\cite{art:broderix,art:grigera02,art:angelani02,art:sciortino03} to identify
the temperature at which  the thermal average of
$n_{im}$  extrapolates to zero
with the critical temperature $T_c$ where the purely dynamical
Mode-Coupling Theory (MCT)
\cite{art:bengtzelius,art:goetze99} predicts a structural arrest of
the liquid. The residual relaxation exhibited by both real and
simulated systems below $T_c$ is to be attributed, within this scenario, to
``activated processes'', i.e. rare transitions over finite energy barriers,
ignored in the mean field approach of MCT.
                                                                                
However, the idea that the slowing down of the liquid around $T_c$  reflects
a geometric transition~\cite{art:grigera02} in the PES has
been recently criticized.
Careful studies have in fact put in doubt the existence of a sudden change in
the sampling of saddles at $T_c$~\cite{art:doliwa03a,art:wales03}
and there are indications that saddles do not disappear completely below $T_c$
\cite{art:doye02,art:fabricius}.
Most worrying, perhaps, is that a saddle-based approach appears unsuitable~\cite{art:berthier03}
for the description of \bem dynamical heterogeneities \eem~\cite{art:kob97,art:yamamoto98,art:kob99},
i.e. rearrangements involving localized subsets of mobile particles,
which have been recognized in the last years as the hallmark of the
supercooled dynamics and whose evidence is hard to glimpse on a 3N-dimensional PES.

In this letter, we analyze the localization properties of the unstable
modes of saddles and quasi-saddles sampled by a supercooled
Lennard-Jones (LJ) mixture. We provide quantitative evidence that the
spatial structure of these unstable modes change from extended to
localized around $T_c$.
We establish for the first time the existence of a direct connection
between saddles and dynamical  heterogeneities in a supercooled liquid.
In fact, the regions of the system
where unstable modes are localized are statistically correlated with
clusters of particles which are highly mobile on the  $\beta$-relaxation timescale.
Our results thus indicate that a crossover in the nature
of the sampled saddles is indeed underlying the emergence of dynamical
heterogeneities and the failure of mean field theories below $T_c$.

Our model is the binary LJ mixture introduced by Wahnstrom
in~\cite{art:wahnstrom}. This mixture has been shown to be a good glass-former
and its properties have been analyzed in-depth
\cite{art:schroeder,art:wahnstrom}, especially as far as dynamical
heterogeneities are concerned~\cite{art:glotzer99}.
Our system is an equimolar mixture of 500 particles
interacting via the LJ potential
$u_{\alpha\beta}(r)=4 \epsilon_{\alpha\beta} [ ( 
    \sigma_{\alpha\beta}/r )^{12} -
  ( \sigma_{\alpha\beta}/r )^6 ]$, with $\alpha,\beta=1,2$ indexes of species, and  
enclosed in a cubic box of side $L=7.2779 \,\sigma_{11}$ with
periodic boundary conditions. Reduced units will be used in the following,
assuming $\sigma_{11}$,  $\epsilon_{11}$, and
$(m_1\sigma_{11}^2/\epsilon_{11})^{1/2}$ respectively  
as unit of distance, energy and
time. The interaction
parameters are $\sigma_{11}=1.0$, $\sigma_{22}=0.833$, $\sigma_{12}=0.917$,
$\epsilon_{11}=\epsilon_{22}=\epsilon_{12}=1.0$. The masses are 
$m_1=1.0$, $m_2=0.5$.
Two different cut-off schemes for the potential have been considered,
cut and shifted (CS) and cut and quadratically shifted (QS) at $r_c=3.0$.
The QS cut-off is obtained adding a term $a+br^2$ with $a$ and $b$
determined to ensure continuity to both $u(r)$ and its first
derivatives at $r_c$. This cut-off scheme has been used to check any
bias in the determination of the properties of the PES due to
discontinuities in the force at the cut-off~\cite{art:shah03}. 
Molecular Dynamics (MD) simulations have been performed in the canonical
ensemble using the Nos\'e-Poincar\'e thermostat~\cite{art:bembenek,art:nose01}.
At the lowest temperature corresponding to an equilibrated supercooled liquid ($T=0.575$),
the length of the run was $2 \times 10^7$ time steps with $\delta
t=0.008$, equivalent to a total simulation time  two orders of magnitude larger than the structural relaxation time.
Our best estimate of the MCT critical temperature, obtained from an analysis of the intermediate scattering functions,  
is $T_c=0.55 \pm 0.01$,
which is slightly lower than the previously reported ones~\cite{art:schroeder}. 

\begin{figure}
\onefigure[scale=0.7]{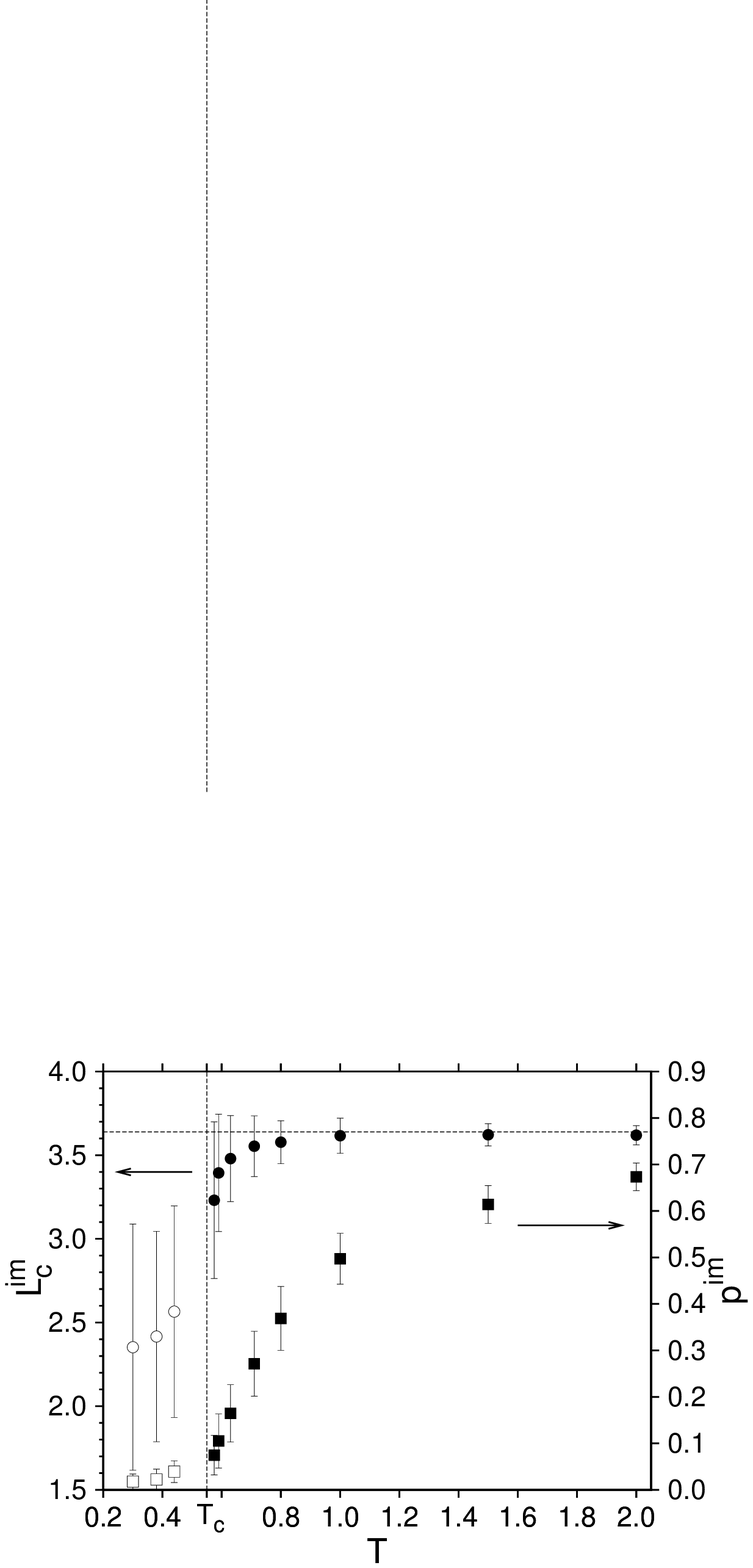}
\caption{Gyration radius $L_c^{im}$ (circles, left axes) and participation ratio
$p^{im}$ (squares, right axes) as a function of $T$, both in the supercooled 
liquid (black) and glass (white). 
Error bars represent one standard deviation of the distribution of
values. The horizontal dashed line
is drawn at $L_{box}/2$.}
\label{fig:loc_sad}
\end{figure}

From each equilibrated run we extracted between 100 and 200 independent
configurations and minimized the mean squared
force $W=1/N\sum_{i} f_i^2$ using the LBFGS
algorithm~\cite{art:nocedal} to locate the nearest saddle. 
It has been recognized~\cite{art:doye02,art:angelani02,art:shah03}
that $W$-minimizations usually stop near
quasi-saddles, i.e. local minima of $W$ where the PES displays one
inflection mode with non-zero force, and that discontinuities in
the cut-off can lead to incomplete minimizations~\cite{art:shah03}.
While other numerical strategies have been proposed for accurately
locating saddles~\cite{art:wales03}, there
has also been evidence~\cite{art:angelani02,art:sciortino03,art:sampoli03} that
quasi-saddles do not differ significantly from true saddles as far as
the influence on the dynamics is concerned. In the
following we will use the term saddle in a wide sense, without
distinction between quasi-saddles and saddles. 
The Hessian matrix at each saddle was diagonalized yielding
the eigenvalues $\nu_{\alpha}^2$ and the eigenvectors
$\eve_i^{\alpha}$, where $\alpha=1,\dots,3N$ is an index of mode and $i=1,\dots,N$ an
index of particle. 
The order $n_{im}$ of a saddle is then defined as the
number of imaginary frequencies in its spectrum.

To have a better understanding of the localization properties of
the unstable modes, we consider the average
\beq
{E_i^{im}}^2 = \frac{1}{n_{im}}\sum_{\alpha=1}^{n_{im}} {\eve_i^{\alpha}}^2
\eeq
of the squared displacements ${\eve_i^{\alpha}}^2$ of atom $i$ over the $n_{im}$
unstable modes.
The vector $E^{im}=(E_1^{im}, \dots, E_N^{im})$
contains an averaged information about the distribution \bem in real
space \eem of the
instabilities associated with the saddle. 
By construction $E^{im}$ is normalized
so that we can define a participation ratio for the unstable
modes of a given saddle in the usual way~\cite{art:bembenek}
\beq
p^{im} = \left( N \sum_{i=1}^N {E_i^{im}}^4 \right)^{-1} .
\eeq
$p^{im}$ will be roughly 1 for extended
instabilities and $O(1/N)$ for a single, localized one.  
Further insight can be gained considering the gyration radius~\cite{art:caprion}
\beq
{L^{im}_c}^2 = \sum_{i=1}^{N} |\rve_i - \rve_g|^2 {E_i^{im}}^2  
\eeq
where $\rve_g = \sum_{i}\rve_i{E_i^{im}}^2$.
For extended instabilities $L^{im}_c \!\approx\! L_{box}/2$. 
Both $p^{im}$ and $L^{im}_c$ refer to a given configuration and quantify
the degree of localization of the unstable modes of 
a saddle as a whole. They will overestimate the size of localized
instabilities when the system is large enough that 
several independent ones are present,
but we will show that this should not be the case for the mixture in consideration.

\begin{figure} 
\onefigure[angle=270,width=1.0\textwidth]{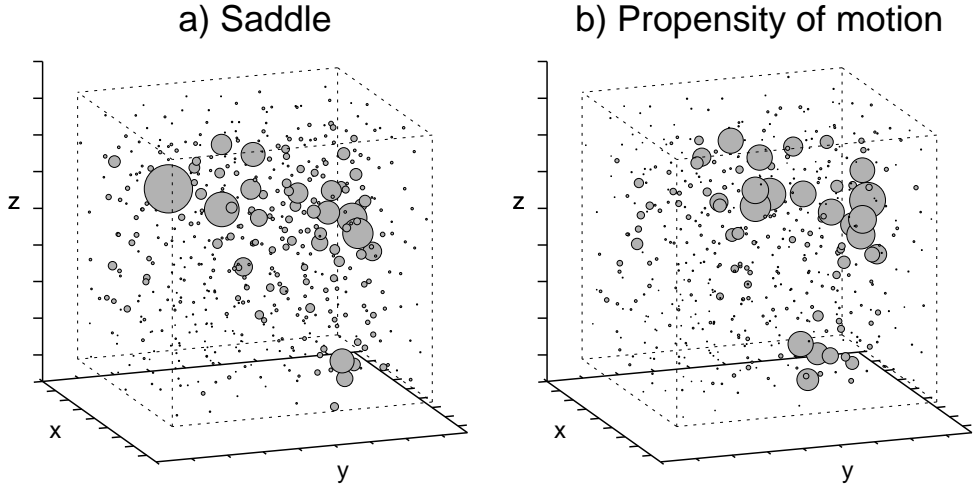}
\caption{Plot a): distribution of average displacements $E_i^{im}$
on the unstable modes of a saddle sampled at $T=0.575$. Plot b): normalized propensity
of motion $E_i^{p}$ of the MD configuration from which the saddle in
plot a) has been located. 
A sphere proportional to $E_i^{im}$ and $E_i^{p}$ is drawn around each
particle, in plot a) and b) respectively.}
\label{fig:snapshot}
\end{figure}

We show in fig.~\ref{fig:loc_sad} the thermal average of the participation ratio $p^{im}$
and gyration radius $L^{im}_c$ as a function of $T$. 
At high $T$ saddles are extended: $p^{im}$ tends to saturate
around 0.7 and $L^{im}_c$ approaches its limiting value of
$L_{box}/2$. In the \bem landscape-influenced \eem range~\cite{art:sastry98},
roughly between $T_c$ and $2T_c$, the instabilities of saddles have a
mixed character. In fact, while $p^{im}$ decrease markedly, 
the gyration radius $L^{im}_c$ still fluctuates
around $L_{box}/2$, indicating that most of the
instabilities of saddles are percolating through the system 
despite their enhanced localization. 
On approaching $T_c$ saddles become
strongly localized and the average $L^{im}_{c}$ decreases abruptly. 
Figure~\ref{fig:loc_sad} thus show that the localized nature of saddles becomes pronounced 
exactly
in the same temperature range where dynamical heterogeneities first
appear~\cite{art:berthier03} and in correspondence to the break-down of
mean-field approaches.

The degree of localization of the instabilities of saddles
can be inspected directly in the snapshot of fig.~\ref{fig:snapshot}a, 
which displays the spatial distribution of ${E_i^{im}}$ in
the simulation box for a saddle sampled at $T=0.575$, slightly above our estimated $T_c$.
A sphere of radius proportional to
${E_i^{im}}$ is drawn around each particle. 
We see that particles involved in the unstable modes, i.e. those
with a large $E_i^{im}$, are strongly clustered. As counterparts, extended
regions of stable particles are observed.
The spatial structure of saddles like that in fig.~\ref{fig:snapshot}a, for which
$n_{im}=7$, casts some new light on a common 
assumption about the nature of saddles sampled by supercooled liquids, namely
that saddles of low order may originate from
non-interacting subsystems, each experiencing a saddle of order
one~\cite{art:sciortino03,art:wales03,art:shell}. 
Figure~\ref{fig:snapshot}a shows that the
7 unstable modes have a strong spatial overlap, being 
concentrated on a rather  compact cluster of
particles. 
Similar features are present in the whole
landscape-influenced range, while
in the hot liquid the fraction of
extended unstable modes increases so that particles are essentially
involved in all of them. 
Thus, the independent subsystems interpretation does not seem to hold 
for saddles sampled by LJ systems in the $T>T_c$ range \footnote{Similar results were in fact obtained
for the BMLJ of~\cite{art:ka1}. We have also checked that this is
the case also for the true saddles we have sampled, for which $W
\approx 10^{-10}$.}.

The emerging scenario seems to mirror the observations of dynamical
heterogeneities~\cite{art:berthier03}. %
The question arises naturally whether there is a direct mapping between
saddles and dynamical heterogeneities, i.e. whether unstable
clusters like that in fig.~\ref{fig:snapshot}a are also the
mobile ones.

To address this issue we followed the approach of~\cite{art:harrowell}, 
who introduced the \bem propensity  of motion \eem of particles as a measure 
of dynamical heterogeneities. 
A configuration is selected from the MD
trajectory at a given temperature $T$. Then several short runs at the same $T$
are 
performed, monitoring the square displacements $\Delta r_i^2(t)$ from
the reference configuration.  
We evaluate the propensity of motion as 
$\lav \Delta r_i^2(t^*) \rav$, where $\lav ... \rav$ 
indicates an average over
independent sets of initial velocities and  $t^*$ corresponds to the maximum of 
the non-Gaussian
parameter $\alpha_2(t)$, which lies in the late 
$\beta$-relaxation regime and is a characteristic timescale
for dynamical heterogeneities~\cite{art:kob97}.  
Using this procedure it is possible to identify the dynamical
heterogeneities that a single configuration will, on average,
give rise to. We have considered up to 500 sets of initial velocities, drawn 
from the appropriate Maxwellian distribution. 
Since the timescales for diffusion are slightly different for the two species
(roughly a factor of 2), we have also tried to choose different values of
$t$ according to the species, but the picture was essentially
unaltered, so we simply used  the value of $t^*$ for the
small particles.

Figure~\ref{fig:snapshot}b shows the distribution in the simulation
box of the normalized propensity of motion 
\beq 
E^p_{i}=\frac{\lav \Delta r_i^2(t^*) \rav}{(\lav \sum_{i=1}^N (\Delta r_i^2(t^*))^2 \rav)^{1/2}}
\eeq 
for the MD configuration whose nearby saddle is actually that of
fig.~\ref{fig:snapshot}a. 
By comparison we see
that the localization and essential morphology of the mobile cluster
identified by the propensity of motion of the MD configuration are well reproduced in the cluster of
average unstable displacements of the nearby saddle.
To assess the statistical relevance of the correlation 
we have analyzed 20 independent
configurations at $T\!=0.575$.
We define as \bem mobile \eem (\bem immobile\eem) those particles for which the
normalized propensity of motion is larger (smaller) than a %
threshold $e^h$ ($e^l$). 
Applying the same cut-off procedure to the average
displacements in the unstable modes $E_i^{im}$, we introduce 
an analogous separation in \bem unstable \eem and \bem stable \eem particles. 
The two threshold values $e^{h}$ and $e^{l}$ could vary or even coincide 
without altering the
overall picture. 
With our choice $e^{l}=0.01, e^{h}=0.045$, 
the fraction of particles belonging on average to each subpopulation
is around 25\%.

\begin{figure}
\onefigure[angle=270,scale=0.4]{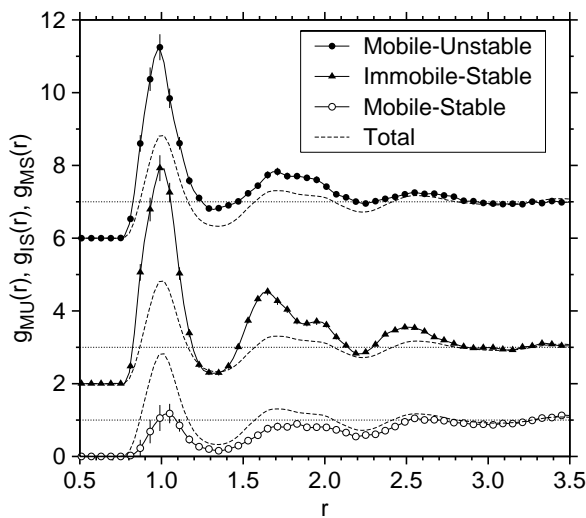}
\caption{From top to bottom: radial distribution functions for 
mobile-unstable (MU), immobile-stable (IS) and mobile-stable (MS)
pairs of particles, compared with the
total distribution function (dashed lines) 
at $T=0.575$ (see text for definitions). Data related to the IS and MU cases have been shifted for clarity. Error bars
represent one standard deviation. }
\label{fig:corr}
\end{figure}

In fig.~\ref{fig:corr}  we show the radial distribution functions at 
$T\!=0.575$ for \bem
mobile-unstable \eem (MU), \bem immobile-stable \eem (IS) and \bem mobile-stable \eem(MS) 
pairs, compared
with the total radial distribution function 
$g(r) = (g_{11}(r)+2g_{12}(r)+g_{22}(r))/4$.
The significant enhancement of the first two coordination shells in $g_{MU}(r)$
and $g_{IS}(r)$ clearly shows that particles with a high (low) propensity of
motion are surrounded, on average, by particles having large (small)
displacements in  the unstable modes of the closest saddle.
Consistently with such  correlation, the $g_{MS}(r)$ (identical to $g_{SM}(r)$) is %
lower than the total $g(r)$. 
Moreover, as expected,  the effect decreases by
increasing temperature.
All this provides statistical  evidence that 
dynamical heterogeneities have spatial correlation
with compact
clusters of particles taking part in the unstable  
modes of nearby saddles.
Interestingly, fig.~\ref{fig:corr} 
also suggests that regions of immobile particles possess distinct structural
properties, as evidenced by the deeper separation between the first and
second shell of neighbours and the enhanced splitting of the second peak in
$g_{IS}(r)$ and in agreement with~\cite{art:kob99}.         

In summary, by performing 
MD simulations for a supercooled LJ mixture we have shown that the
relaxation channels associated with the unstable modes of saddles and
quasi-saddles sampled along the MD trajectory crossover from
spatially extended to localized around $T_c$. 
These localized unstable modes display
non-trivial spatial correlations
with the dynamical
heterogeneities identified by the propensity of
motion~\cite{art:harrowell} in the $\beta$-relaxation timescale.
The novel finding of such a
correlation may represent a step ahead in the
understanding of the heterogeneous nature of the supercooled
dynamics. Theoretical and computational tools that have been developed
for the study of the PES~\cite{buk:wales,art:cavagna03}, may now reveal
their utility in a direct analysis of dynamical heterogeneities. 
We note that other observables, e.g. free volume, have been
recently addressed as possible origins of dynamical heterogeneities, 
but their spatial
correlation with local mobility has been either found to be poor 
\cite{art:widmercooper05} or investigated below the glass transition 
\cite{art:teichler06}.
In this perspective, our results point to a key role of
saddles and quasi-saddles in bridging the gap between
the PES description of supercooled liquids and approaches which 
focus more directly on the dynamics 
like the
heterogeneity-based picture proposed by Berthier \etal \cite{art:berthier03} 
or the mean field MCT.

\acknowledgments
We would like to thank F. Sciortino for useful comments.

\end{document}